\documentclass[%
 reprint,
 amsmath,amssymb,
 aps,
]{revtex4-2}

\usepackage{graphicx}
\usepackage{subfig}
\usepackage{dcolumn}
\usepackage{bm}
\usepackage{braket}
\usepackage{amsmath}
\usepackage[colorlinks=true, linkcolor=blue, citecolor=blue, urlcolor=blue]{hyperref}

\begin{document}

\preprint{APS/123-QED}

\title{Entangled Quantum Walkers for Secure Quantum Key Distribution}

\author{Chia-Tso Lai}
 \affiliation{Independent Researcher, Leipzig, Germany}
 \email{chiatsolai@gmail.com}

\date{\today}

\begin{abstract}
Quantum Key Distribution (QKD) is an emerging cryptographic method designed for secure key sharing. Its security is theoretically guaranteed by fundamental principles of quantum mechanics, making it a leading candidate for future communication protocols. Quantum Random Walks (QRWs), on the other hand, are quantum processes that exhibit intriguing phenomena such as interference and superposition, enabling the generation of decentralized and asymmetric probability distributions. Inspired by both fields of study, we propose a novel QKD protocol based on two entangled quantum walkers. Our protocol exploits the unique correlations between the walkers at extremal positions of the walk to establish secret keys shared exclusively by the two parties. The security of the protocol is augmented by analyzing the joint probability distributions of the walkers’ measured positions and their associated coin states.
\end{abstract}

\maketitle



\section{\label{sec:Introduction}Introduction}
\begin{figure}
    \centering
    \includegraphics[width=0.75\linewidth]{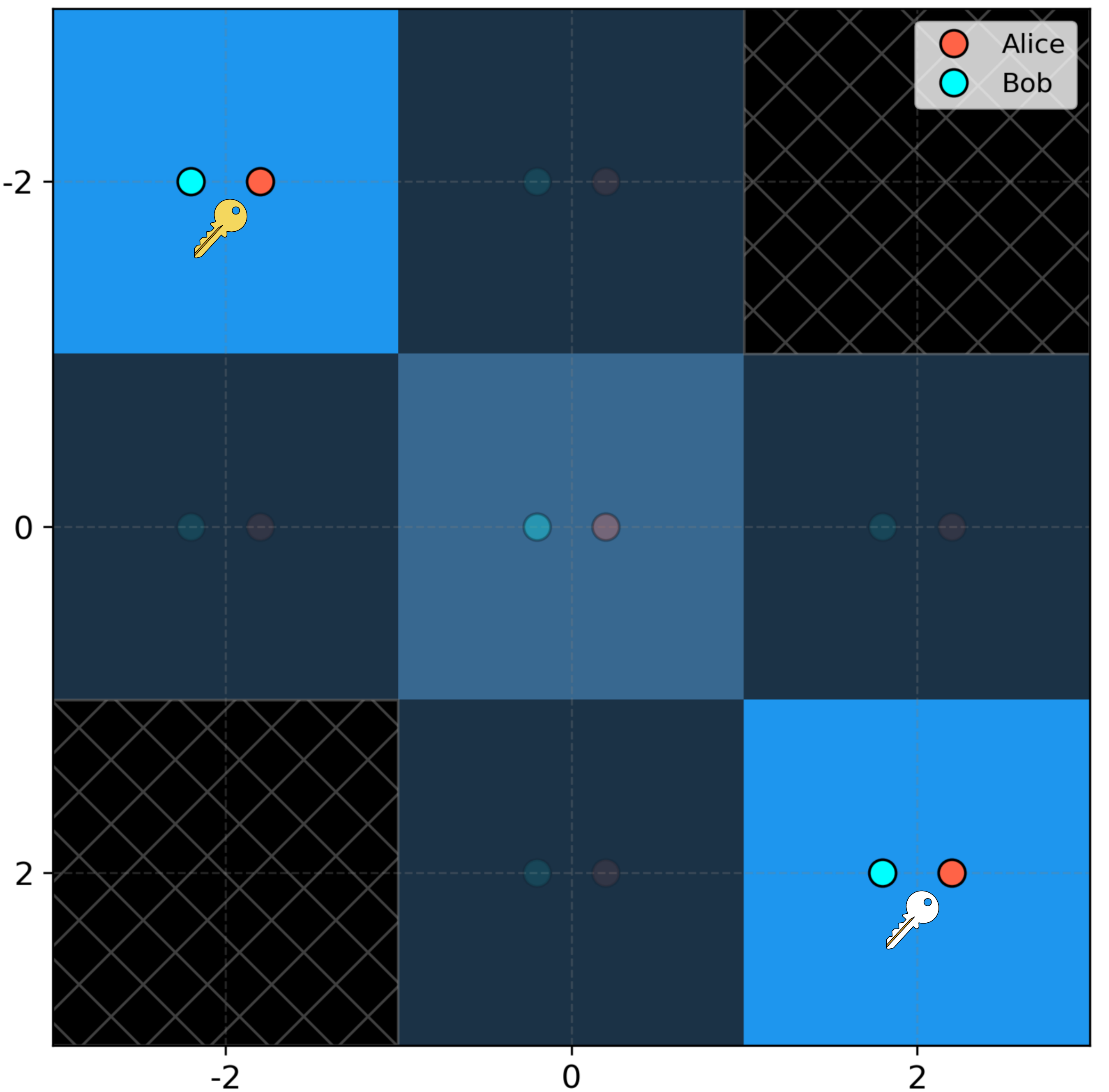}
    \caption{Entangled walkers Alice and Bob obtain shared secret keys at opposite corners of the joint quantum walk.}
    \label{fig:entangled_walker_schematic}
\end{figure}
Quantum Key Distribution (QKD) is a robust cryptographic scheme devised to defend against eavesdropping during secret communication. It leverages quantum properties such as the no-cloning theorem and quantum entanglement to ensure security. A variety of QKD protocols have been proposed since its inception. For instance, the pioneering BB84 protocol \cite{bennett1984proceedings} introduced the first example of a prepare-and-measure QKD scheme. Any attempt at observation by an eavesdropper disturbs the quantum state if the measurement basis differs from the preparation basis, thereby revealing the intrusion. The E91 protocol \cite{ekert1991quantum} laid the foundation for entanglement-based QKD. E91 uses Bell’s theorem as a security guarantee against eavesdropping. The strength of the correlations between the measured keys can be quantified using the CHSH test \cite{clauser1969proposed}, which signals a potential attack if the correlations fall below a certain classical limit. Our proposed protocol can be categorized as a prepare-and-double-measure QKD scheme that incorporates entanglement to achieve secure and exclusive key distribution. We begin the protocol with an entangled "coin" pair prepared by the sender and shared with the receiver. Each coin is then entangled with a "walker" at the respective end of the communication channel. A specific connection between the walkers is established via entanglement swapping \cite{bennett1993teleporting}, in which a Bell state measurement (BSM) is performed on the coin pair once it is reunited at the sender's end. This process creates an entangled state shared between the two walkers, enabling the parties to derive a pair of secret keys (Fig.~\ref{fig:entangled_walker_schematic}). The security of our protocol relies on quantum entanglement, though the verification method differs from the standard CHSH formulation. \\

Quantum Random Walks (QRWs) \cite{aharonov1993quantum} are the quantum analog of classical random walks, which form the basis of many stochastic algorithms. Due to quantum interference and superposition, QRWs exhibit markedly different behavior, often resulting in faster, ballistic spreading of probability distributions compared to the centralized, diffusive binomial distributions of classical walks (Fig.~\ref{fig:qrw_dist}). In this work, we focus on discrete coined quantum random walks \cite{brun2003quantum}, where the evolution of the walker is governed by a quantum coin state and a unitary coin-flip operator. Motivated by the properties of coined QRWs, we investigate the behavior of joint QRWs involving two entangled walkers. These joint walks show intriguing correlations in position space, particularly at the edges of the distribution. We demonstrate that these spatial correlations allow both parties to share mutual information about the measured positions, without revealing the actual outcomes. As a result, the walkers’ positions can be effectively used to encode secret keys, making entangled QRWs a promising building block for secure quantum communication protocols.

\section{\label{sec:Theory}Theory}

\subsection{\label{sec:qrw} Discrete Coined QRW}

A discrete coined QRW is composed of several elements: the number of steps $s \in \mathbb{Z}^+$, the walker's position state $\ket{x}$ (with $x \in \mathbb{Z}$), the coin state $\ket{c}$, a unitary coin-flip operator $\hat{U}(\theta,\lambda)$, and a shift operator $\hat{S}$. The entire system can be described by the product state $\ket{\psi} = \ket{c} \otimes \ket{x}$. The QRW evolves by repeatedly applying the coin-flip operator to the coin, followed by the shift operator applied to the position register, conditioned on the coin state.

We consider a general coin-flip operator $\hat{U}(\theta,\lambda)$ given by:
\[
\hat{U}(\theta,\lambda) = \begin{bmatrix}
\cos(\frac{\theta}{2}) & -e^{i\lambda}\sin(\frac{\theta}{2}) \\
\sin(\frac{\theta}{2}) & e^{i\lambda}\cos(\frac{\theta}{2})
\end{bmatrix}
\]
The conditional shift operator $\hat{S}$ is defined as:
\begin{equation}
    \hat{S} = \ket{0}\bra{0} \otimes \sum_{i}\ket{i-1}\bra{i}+ \ket{1}\bra{1} \otimes \sum_{i}\ket{i+1}\bra{i}
\end{equation}

Assume the walker starts from an initial position $\ket{x_0}$ and the coin is prepared in a state $\ket{c_0} = k_0\ket{0}+k_1\ket{1}$, with $k_1 = e^{i\phi}\sqrt{1-k_0^{2}}$, resulting in the initial product state $\ket{\psi^0} = \ket{c_0}\ket{x_0}$. After one step of QRW, the state becomes:
\begin{equation}
    \begin{split}
    \ket{\psi^1} &= \hat{S}(\hat{U} \otimes \hat{I})\ket{\psi^0} \\ &= \ket{0}\otimes(k_0\cos(\frac{\theta}{2})-k_1e^{i\lambda}\sin(\frac{\theta}{2}))\ket{x_0-1} \\ & \quad + \ket{1}\otimes(k_0\sin(\frac{\theta}{2})+k_1e^{i\lambda}\cos(\frac{\theta}{2}))\ket{x_0+1}
    \end{split}
\end{equation}
After $s$ steps, the state evolves to:
\begin{equation}
    \ket{\psi^s} = \ket{0}\otimes\sum_{i=-s}^{s-2}A_{i}^{(s)}\ket{x_0+i}+\ket{1}\otimes\sum_{j=-s+2}^{s}B_{j}^{(s)}\ket{x_0+j}
\end{equation}
where the amplitudes $A_{i}^{(s)}$ and $B_{j}^{(s)}$ vanish at positions where the shift from $-s$ or $s$ is odd (i.e., $i,j = -s+1, -s+3, \dots, s-1$ ). The amplitudes can be computed recursively by:
\[
\begin{cases}
A_i^{(s+1)} =  A_{i+1}^{(s)} \cdot \cos(\frac{\theta}{2}) - B_{i+1}^{(s)} \cdot e^{i\lambda}\sin(\frac{\theta}{2}) \\
B_i^{(s+1)} = A_{i-1}^{(s)} \cdot \sin(\frac{\theta}{2}) + B_{i-1}^{(s)} \cdot e^{i\lambda}\cos(\frac{\theta}{2})
\end{cases}
\]
with initial conditions $A_i^{(0)} = k_0\delta_{i0}$ and $B_i^{(0)} = k_1\delta_{i0}$.

\begin{figure}
    \centering
    \includegraphics[width=0.9\linewidth]{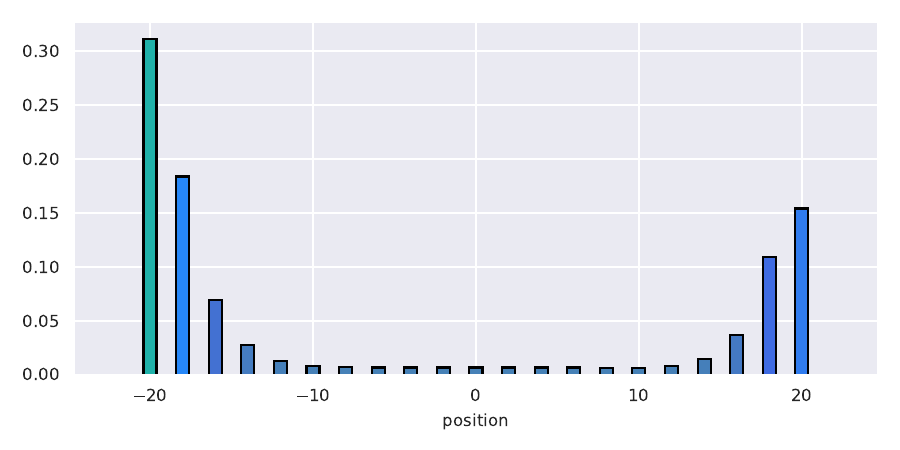}
    \caption{Probability distribution after 20 steps of a discrete coined QRW.}
    \label{fig:qrw_dist}
\end{figure}

\subsection{\label{sec:entangeled qrw} Entangled Quantum Walks}

In our proposed QKD protocol, two QRWs are implemented with a shared entangled coin pair $\ket{c_A,c_B} = k_0\ket{00}+k_1\ket{11}$. Both walkers start at the origin, $x_0 = y_0 = 0$, and evolve for the same number of steps using publicly agreed coin-flip operators, $\hat{U}_A$ and $\hat{U}_B$, respectively. This results in a joint position distribution over the basis states $\ket{x}_A \ket{y}_B$ at the end of the walk. The state after $s$ steps is given by:
\begin{equation}\label{eq:entangled_qrw_state}
\begin{split}
    \ket{\psi^s} &= \ket{00}\otimes\sum_{x=-s}^{s-2}\sum_{y=-s}^{s-2}A_{xy}^{(s)}\ket{x}_A\ket{y}_B \\ & \quad +\ket{01}\otimes\sum_{x=-s}^{s-2}\sum_{y=-s+2}^{s}B_{xy}^{(s)}\ket{x}_A\ket{y}_B \\ & \quad +\ket{10}\otimes\sum_{x=-s+2}^{s}\sum_{y=-s}^{s-2}C_{xy}^{(s)}\ket{x}_A\ket{y}_B \\ & \quad +\ket{11}\otimes\sum_{x=-s+2}^{s}\sum_{y=-s+2}^{s}D_{xy}^{(s)}\ket{x}_A\ket{y}_B 
\end{split}
\end{equation}
where the amplitudes $A_{xy}^{(s)}$, $B_{xy}^{(s)}$, $C_{xy}^{(s)}$ and $D_{xy}^{(s)}$ again vanish at positions where the shift from $-s$ or $s$ is odd. Similar to a single-walker QRW, these amplitudes can be computed recursively. First we transform the amplitudes for the same positions $(x,y)$ with the composite coin-flip operator:
\begin{equation}
    \begin{bmatrix}
        A_{xy}^{\prime (s)} \\
        B_{xy}^{\prime (s)} \\
        C_{xy}^{\prime (s)} \\
        D_{xy}^{\prime (s)} 
    \end{bmatrix}
    =
    (\hat{U}_A \otimes \hat{U}_B)
    \begin{bmatrix}
        A_{xy}^{(s)} \\
        B_{xy}^{(s)} \\
        C_{xy}^{(s)} \\
        D_{xy}^{(s)}
    \end{bmatrix}
\end{equation}
Then, the recursive relations can be formulated as:
\begin{equation}
    \begin{cases}
        A_{xy}^{(s+1)} = A_{x+1, y+1}^{\prime (s)}\\
        B_{xy}^{(s+1)} = B_{x+1, y-1}^{\prime (s)}\\
        C_{xy}^{(s+1)} = C_{x-1, y+1}^{\prime (s)}\\
        D_{xy}^{(s+1)} = D_{x-1, y-1}^{\prime (s)}\\
    \end{cases}
\end{equation}
with initial conditions $A_{xy}^{(0)} = k_0\delta_{x0}\delta_{y0}$, $B_{xy}^{(0)} = 0$, $C_{xy}^{(0)} = 0$, and $D_{xy}^{(0)} = k_1\delta_{x0}\delta_{y0}$.

We proceed to implement entanglement swapping on the state $\ket{\psi^s}$ to establish correlations between the two walkers' position distributions. This is achieved by performing a Bell state measurement (BSM) on the entangled coin pair $\ket{c_A, c_B}$.  The BSM is realized by applying a CNOT gate to the pair, followed by a Hadamard gate on the control coin qubit, resulting in the state $\ket{\psi^{\prime s}}$ prior to measuring the coin pair. For notational convenience, we define the summations in Eq.~(\ref{eq:entangled_qrw_state}) as:
\begin{equation}
    \begin{split}
        \textbf{A}^{(s)} &= \sum_{x=-s}^{s-2}\sum_{y=-s}^{s-2}A_{xy}^{(s)}\ket{x}_A\ket{y}_B\\
        \textbf{B}^{(s)} &= \sum_{x=-s}^{s-2}\sum_{y=-s+2}^{s}B_{xy}^{(s)}\ket{x}_A\ket{y}_B\\
        \textbf{C}^{(s)} &= \sum_{x=-s+2}^{s}\sum_{y=-s}^{s-2}C_{xy}^{(s)}\ket{x}_A\ket{y}_B\\
        \textbf{D}^{(s)} &= \sum_{x=-s+2}^{s}\sum_{y=-s+2}^{s}D_{xy}^{(s)}\ket{x}_A\ket{y}_B\\
    \end{split}
\end{equation}
This allows us to express Eq.~(\ref{eq:entangled_qrw_state}) more compactly such that the transformed state $\ket{\psi^{\prime s}}$ becomes:
\begin{equation}\label{eq:entangled_walkers}
    \begin{split}
        \ket{\psi^{\prime s}} &= \frac{1}{\sqrt{2}}\ket{00} \otimes \left[\textbf{A}^{(s)} + \textbf{D}^{(s)}\right] \\ & \quad + \frac{1}{\sqrt{2}}\ket{10} \otimes \left[\textbf{A}^{(s)} - \textbf{D}^{(s)}\right] \\ & \quad + \frac{1}{\sqrt{2}}\ket{01} \otimes \left[\textbf{B}^{(s)} + \textbf{C}^{(s)} \right] \\ & \quad + \frac{1}{\sqrt{2}}\ket{11} \otimes \left[\textbf{B}^{(s)} - \textbf{C}^{(s)} \right]
    \end{split}
\end{equation}
The Bell measurement collapses the state, projecting the coin pair onto one of the four possible outcomes: ${\ket{00}, \ket{01}, \ket{10}, \ket{11}}$.

We observe that the expansion of $\textbf{A}^{(s)} + \textbf{D}^{(s)}$ (and likewise $\textbf{A}^{(s)} - \textbf{D}^{(s)}$) contains the terms $A_{-s,-s}^{(s)}\ket{-s}_A\ket{-s}_B$ and $D_{s,s}^{(s)}\ket{s}_A\ket{s}_B$, while the terms $\ket{-s}_A\ket{s}_B$ and $\ket{s}_A\ket{-s}_B$ are absent. This implies that if the BSM outcome is $\ket{00}$ (or $\ket{10}$), the probability of the two walkers being on opposite ends of the distribution is zero (see Fig.~(\ref{fig:bsm00})). In other words, if both walkers are found at the extremities of their distributions, they must be at the same extremity (either both at $-s$ or both at $s$). A similar argument holds for the expansions of $\textbf{B}^{(s)} + \textbf{C}^{(s)}$ and $\textbf{B}^{(s)} - \textbf{C}^{(s)}$. In these cases, if the measurement outcome is $\ket{01}$ or $\ket{11}$ and both walkers are at their extremities, then they cannot be located at the same end of the distribution—only opposite ends are possible in such a scenario (see Fig.~(\ref{fig:bsm01})).

\section{\label{sec:protocol}QKD via Entangled Quantum Walkers}

Given the exclusive correlations between two entangled quantum walkers, as described by Eq.~(\ref{eq:entangled_walkers}), a secure QKD protocol can be designed by utilizing the two extremities of the position distribution as secret keys. Based on this idea, we propose a robust quantum communication scheme between two parties—Alice and Bob—consisting of the following steps:
\begin{enumerate}
    \item \textbf{Entangled coin pair preparation}: Alice chooses an initial coin state $\ket{c_A} = k_0\ket{0} + k_1\ket{1}$, with which she prepares an entangled coin pair $\ket{c_A, c_B} = k_0\ket{00} + k_1\ket{11}$. The coin state parameters are only known to Alice.
    
    \item \textbf{Transmission}: Alice sends one qubit from the entangled coin pair to Bob via a quantum channel.
    
    \item \textbf{Distribution of QRW parameters}: Alice and Bob communicate over a classical channel to agree on the parameters of the QRWs, including the coin-flip operators $\hat{U}_A(\theta_1,\lambda_1)$ and $\hat{U}_B(\theta_2,\lambda_2)$, and the number of steps $s$. Choosing smaller phase angles $\theta_1$ and $\theta_2$ increases the probability at the extremities of the position distribution (see Fig.~\ref{fig:qrw_dist}), thereby improving the likelihood of successful key generation.
    
    \item \textbf{QRW implementation}: Alice and Bob independently implement the QRW using the agreed-upon parameters.
    
    \item \textbf{Entanglement swapping}: Bob sends his coin qubit back to Alice, who then performs a BSM to transfer the entanglement to the walkers’ position states. The BSM outcome is kept by Alice as a reference for key inference.
    
    \item \textbf{Position measurement}: Both parties measure the positions of their walkers, obtaining values $a, b \in \{-s, -s+2, \ldots, s\}$.
    
    \item \textbf{Announcement}: After multiple rounds of steps 1 to 6, Alice and Bob announce the rounds in which their walkers were found at extremal positions (without revealing which extremity). In rounds where either Alice or Bob did not find their walker at an extremal position, Bob discloses the measured position to Alice.
    
    \item \textbf{Key generation}: Using the BSM outcomes and her own measurement results, Alice infers which position Bob measured in the rounds where both observed extremal values. Bob’s measured positions ($\pm s$) are then used as the sifted keys.
    
    \item \textbf{Security check}: The protocol includes two layers of security verification. First, Alice checks whether the BSM outcomes follow the expected probability distribution of the coin pair, allowing a deviation up to $\epsilon_c$ based on a predefined metric. Second, the joint distribution of measured positions is compared against the theoretical model. If either test fails, the protocol is aborted.
\end{enumerate}
 
\section{\label{sec:example}2-step QRW protocol}

\begin{figure}
    \centering
    \includegraphics[width=0.9\linewidth]{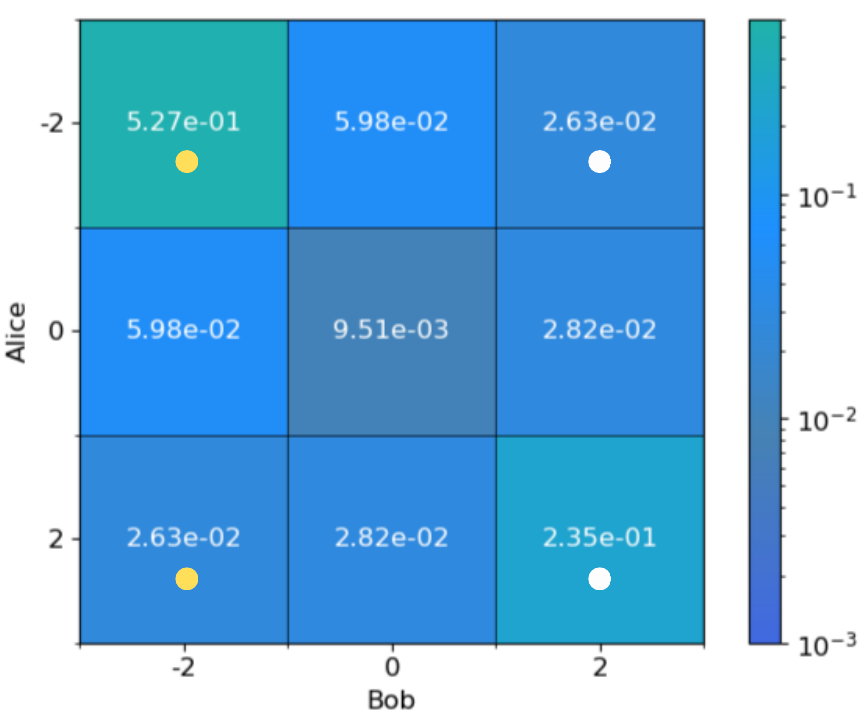}
    \caption{Joint probability distribution of Alice and Bob's positions, considering all BSM outcomes. The four corners of the distribution correspond to scenarios where key pairs can be generated. Yellow dots denote the shared key value -2, while white dots indicate the key value 2.}
    \label{fig:two_qrw_all_prob}
\end{figure}

\begin{figure}[htbp]
    \centering
    \subfloat[$\ket{c_A,c_B}=\ket{00}$]{\label{fig:bsm00}
        \includegraphics[width=0.9\columnwidth]{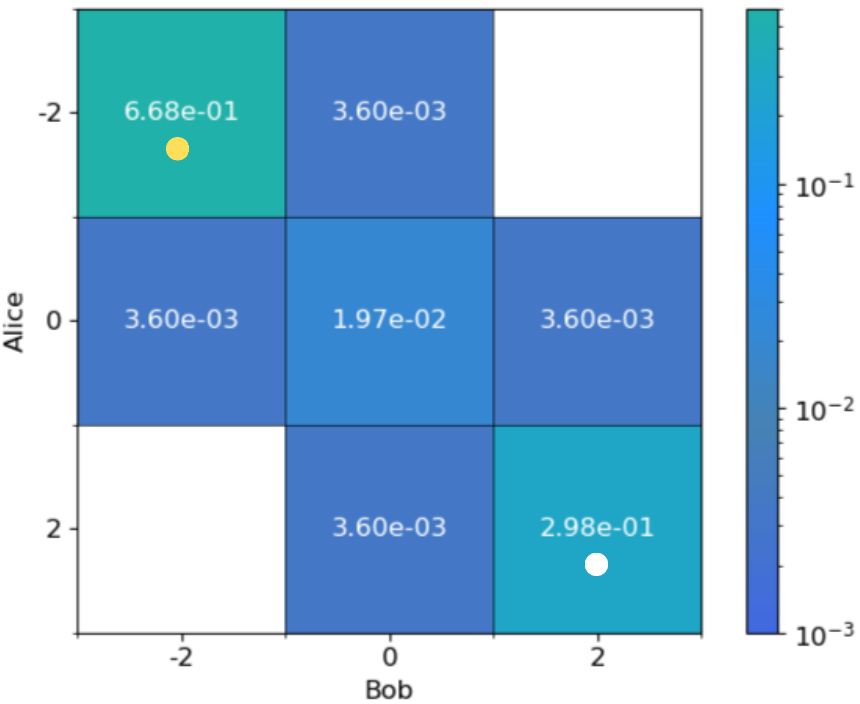}
    }
    \hfill 
    \subfloat[$\ket{c_A,c_B}=\ket{01}$]{\label{fig:bsm01}
        \includegraphics[width=0.9\columnwidth]{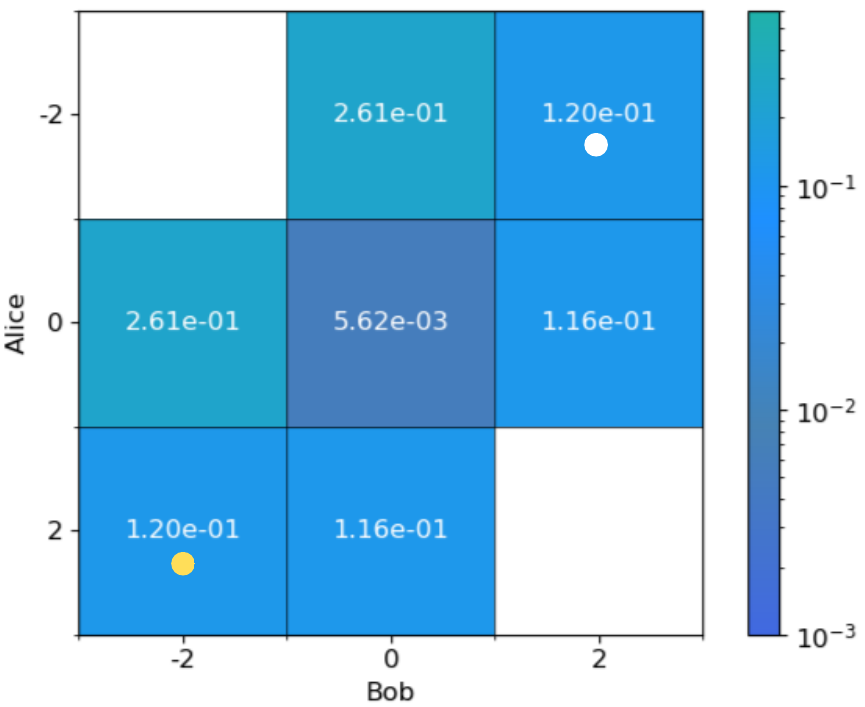}
    }
    \caption{Joint probability distribution of measured positions when the BSM outcome is (a) $\ket{c_A,c_B} = \ket{00}$ and (b) $\ket{c_A,c_B} = \ket{01}$. Empty grid cells indicate zero probability for the corresponding position pair. The parity of the BSM outcome determines the correlation pattern of the two walkers, allowing Alice to infer Bob's key. \label{fig:qrw_bsm}}
\end{figure}

In this section, we demonstrate a QKD protocol based on 2-step QRWs, i.e., with $s = 2$. The 2-step setup strikes a balance between hardware feasibility and complexity, making it suitable for near-term implementation. It requires only four qubits in total for the coin and position registers per party, which aligns well with current hardware limitations. Furthermore, the 2-step protocol is statistically efficient for security verification, as it yields only nine possible outcomes in the joint position distribution. This small outcome space reduces the number of required protocol rounds to obtain a meaningful match between the observed and theoretical distributions. While a 1-step QRW protocol is also viable for practical realization, its limited structure may not adequately demonstrate the features of generic $s$-step schemes. In contrast, the increased complexity of the 2-step QRW allows for a more representative example, and potentially offers stronger security guarantees against eavesdropping, due to the higher-dimensional correlations available for verification.

Assume Alice and Bob choose the same coin-flip operator, $\hat{U}_A(\theta,\lambda) = \hat{U}_B(\theta,\lambda)$, to perform the 2-step QRW with the following parameters: $\theta = 0.635$, $\lambda = 0$, and initial coin state $\ket{c_A} = 0.85\ket{0} + 0.527 e^{i\frac{\pi}{4}} \ket{1}$. The joint probability distribution of all possible measured positions, aggregated over all BSM outcomes, is summarized in Fig.~\ref{fig:two_qrw_all_prob}. The four corners of the heatmap correspond to measurement outcomes that enable key generation. In this instance, a valid key pair can be generated with a probability of approximately 81\%. For 2-step QRW protocols, the sifted key fraction is calculated to be $\cos^{4}(\frac{\theta}{2})$. Other measurement results, while not used for key generation, serve as references for security verification. Fig.~\ref{fig:bsm00} shows the joint distribution of Alice’s and Bob’s positions when the BSM outcome is $\ket{00}$, which occurs with probability 39\%. The heatmap reveals that when Bob measures $-2$ (respectively, $2$), Alice has zero probability of measuring $2$ (respectively, $-2$), ensuring the exclusivity of the shared key. Similarly, Fig.~\ref{fig:bsm01} shows the distribution when $\ket{c_A,c_B} = \ket{01}$, which occurs with probability 11\%. In this case, when Bob measures $-2$ ($2$), Alice has zero probability of measuring $-2$ ($2$), maintaining exclusivity under a different correlation pattern.

\section{\label{sec:conclusion}Conclusion and Outlook}

We have shown that the unique correlations between two entangled quantum walkers, established via entanglement swapping, can be harnessed as a secure resource for quantum communication, enabling the development of a powerful QKD protocol. The behavior of the walkers at their spatial extremities, together with the Bell state measurement outcomes, provides both parties with exclusive information about the position states. Furthermore, the BSM results and the joint position distributions of the QRWs serve as a foundation for security verification, enhancing the robustness of the protocol. We also demonstrated that a 2-step QRW protocol can achieve a high sifted key fraction with a proper choice of parameter $\theta$ in the coin-flip operator.

For future research, the impact of noisy quantum channels and specialized attacks on this new protocol should be a primary focus. Additionally, incorporating CHSH tests or other security verification methods into the protocol may offer valuable enhancements. Experimental realizations, alongside suitable error reconciliation and privacy amplification schemes, will be essential for evaluating the practical viability of the proposed protocol. Finally, we emphasize the innovative and interdisciplinary nature of this work: the integration of entangled quantum random walks into quantum cryptography not only showcases a novel application of QRWs but also enriches the landscape of QKD protocols. 

\section*{Code Availability}

All simulation code for the entangled quantum random walks and the QKD protocol is openly available at: \href{https://github.com/jordan0809/Entangled-Walkers-QKD}{Entangled-Walkers-QKD}


\nocite{*}

\bibliography{main}

\end{document}